\documentclass[a4paper]{article}
\usepackage{axodraw}
\usepackage{mymacros}
\usepackage{epsfig}
\title{Singular Cross Sections in Muon Colliders}
\author{Chris Dams\footnote{chrisd@sci.kun.nl}, Ronald Kleiss\footnote{kleiss@sci.kun.nl}\\\small University of Nijmegen, The Netherlands}
\date{December 19, 2002}
\def\th{\textsl{th}}
\def\centerpict#{\vcenter\bgroup\hbox\bgroup\aftergroup\egroup\let\centerpict}
\begin{document}

\maketitle

\begin{abstract}
We address the problem that the cross section for the collisions of unstable
particles diverges, if calculated by standard methods. This problem is
considered for beams much smaller than the decay length of the unstable
particle, much larger than the decay length and finally also for pancake-%
shaped beams. We find that in all cases this problem can be solved by
taking into account the production/propagation of the unstable particle
and/or the width of the incoming wave packets in momentum space.
\end{abstract}

\section{Introduction}
When one applies the Feynman Rules and the Golden Rule to a collision
of unstable particles, the cross section turns out to diverge.
We summarize how this happens. The
divergence occurs for instance in the graph
\beq\label{eq:fmdiag}
\centerpict{
\begin{picture}(150,130)
\ArrowLine(10,10)(50,40)\Text(30,25)[rb]{$\mu^-(p_1)$}
\ArrowLine(50,40)(100,20)\Text(77,31)[lb]{$W^-$}
\ArrowLine(100,20)(130,5)\Text(133,5)[l]{$e^-(q_1)$}
\ArrowLine(130,35)(100,20)\Text(133,35)[l]{$\bar\nu_e(q_2)$}
\ArrowLine(50,40)(50,90)\Text(55,65)[l]{$\nu_\mu(k)$}
\ArrowLine(50,90)(10,130)\Text(30,110)[rt]{$\mu^+(p_2)$}
\ArrowLine(90,130)(50,90)\Text(70,110)[lt]{$W^+(q_3)$}
\end{picture}}.
\eeq
The lower half of this diagram looks like the decay of the muon. The
consequence is that the momentum~$k$ may be on its mass shell. After all,
that is what one gets from the decay of a muon: a muon neutrino on its
mass shell.
The propagator of this muon neutrino contains a factor
\beq\label{eq:prop}
\Delta(k)=\frac i{k^2+i\epsilon}
\eeq
that gets squared and integrated over as the Golden Rule tells us to do.
Because the neutrino that contributes the just given factor can be on
its mass shell, we find that
in lowest order in~$\epsilon$ the result will go as~$\epsilon^{-1}$.
This is a divergent quantity since $\epsilon$ is taken to be
infinitesimal. This problem is of relevance for muon colliders as was
already noted in~\cite{problem}.

The question that naturally poses itself is if the cross section really
diverges,
and if so, how this can be regularized, and if not, how we can compute it.
In general we can say that there are two possibilities to solve the question
of the divergent cross section. The first is that one takes into account that
unstable particles cannot really be in or out states. This can be done by
considering Feynman graphs that include the production process
of the unstable particle. The
second possibility is that one takes into account that the incoming
wave packets are not really sharp momentum states but that we always have
interference between states with the same total momentum but with some
of the momentum moved from one incoming particle to the other. This is
of importance if the peak structure of the matrix element is sharper than
the size of the incoming wave packet in momentum space. We indeed have this
in the case of the unstable particle, since the propagator that causes
the divergence has no particular size attribute, except for
the~$\epsilon$, which is taken to be infinitesimal.

This problem has mainly been discussed for the case that would be applicable
to, for instance, muon colliders. Cf.,~\cite{inter}, \cite{phys},~\cite{sing}.
We will mostly address a case of academic interest that has been
considered alongside the realistic case, namely that of very wide beams
(i.e., much larger than the decay length of the muon).
We will try to apply the methods applicable to the realistic case also in
the case of infinitely wide beams. For this we will need some modifications
to the realistic case. This case is considered
in the next section. We reconsider it because we use different, more covariant
notations, best introduced in a somewhat more familiar setting,
and also because we have some points to add.

Finally we consider the case of pancake-shaped beams (i.e., much larger than
the decay length in the transversal directions and much smaller than that in
the longitudinal direction).

\section{The Realistic Case}
\label{sec:realistic}
Here we discuss the
solution as would be applicable in muon colliders, that is, for beams of
which the size
is much smaller than the decay length of the unstable particle.
In this case the propagator of the unstable particle
confines its momentum to the mass-shell and because production and collision
can be chosen to be well-separated in space, we need not worry about the
production process. There is, however, still the point that the matrix
element is peaked sharply enough to notice interference
between states where some momentum is shifted from one incoming particle
to the other. This case has been solved in~\cite{sing}, however we
again present this here using notations that are manifestly covariant.

The quantum distribution function~$n(p,r)$, introduced in~\cite{inter}, is used
to describe the particles.
It is defined to be given by
\beq
n(p,r)=\frac m{(2\pi)^3}\int d^4\Delta\,\delta(p\cdot\Delta)\,
        \phi(p+\tfrac12\Delta)\,\phi(p-\tfrac12\Delta)^*\,
        e^{-i\Delta\cdot r}.
\eeq
The $\delta(p\cdot\Delta)$ is used to confine
the particle to its mass shell. This implies that we use the approximation
that the components of $\Delta$ are much smaller than the ones of~$p$.
The quantum distribution function contains
as much information about the state of a particle
as a density matrix. 
From it the probability densities in momentum and position space can be found.
They are given by
\beq
\rho(p)=\frac{p^0}m\int d^3r\,n(p,r);\qquad
	\rho(r)=\int\tie{p} \frac{p^0}m n(p,r).
\eeq
Both are (in the approximation that they are sharply peaked in momentum
space) zeroth components of four-vectors. This should, of course, be the case
with densities.
The probability measure that belongs to them is
respectively~$\tie{p}$ (which is by definition equal to~%
$\frac{d^3p}{(2\pi)^32p^0}$) and~$d^3r$.
This looks pretty much as if $n(p,r)$ were a joint
probability for $r$~and $p$, but of course such a thing cannot really
exist in quantum mechanics and actually~$n(p,r)$
does not need to be a real function, which discards it as a probability
density.

We extend the definition of the luminosity to
\beq\label{eq:stablum}
dL(p_1,p_2,\rho)=\tie{p_1}\,\tie{p_2}\frac{\sqrt{-\mathrm{GD}}}{m_1m_2}
                \int d^4r\,n_1(p_1,r)n_2(p_2,r+\rho).
\eeq
This reduces to the normal definition for $dL$ by setting~$\rho=0$.
The GD that occurs here stands for ``Gramm determinant'' and is equal to
$p_1^2p_2^2-(p_1\cdot p_2)^2$.
Also the definition of the cross section is extended, namely to
\beq
\eqalign{
d\sigma(p_1,p_2,\Delta)&=
        \frac1{4\sqrt{-\mathrm{GD}}}
        (2\pi)^4\delta(p_1+p_2-q)\,\tie{q_1}\cdots\tie{q_n}\cr
        &\qquad\Mel(p_1+\tfrac12\Delta,p_2-\tfrac12\Delta,q)
        \Mel(p_1-\tfrac12\Delta,p_2+\tfrac12\Delta,q)^*.\cr
}
\eeq
This gives the well-known definition for the cross section if we
take~$\Delta=0$. The exact expression for the number of events~$W$
(see for instance~\cite{ryder}) can, using these definitions, be written as
\beq\label{eq:numevents}
W=\int dL(p_1,p_2,\rho)\frac{d^4\Delta}{(2\pi)^4}d^4\rho\,e^{-i\Delta\cdot\rho}
        d\sigma(p_1,p_2,\Delta).
\eeq
Normally one assumes that $d\sigma$ does not vary much with~$\Delta$ and
for that reason it is safe to put $\Delta=0$.
Then the integrals over $\Delta$~and
$\rho$ become trivial and the familiar result that ``number of events is
cross section times luminosity'' is obtained. In our case, however,
$\sigma$ has a pole at $\Delta=0$ so this approximation cannot be made.

The reasoning about how to get
a finite cross section is essentially identical
to what is given in~\cite{sing},
so we will not present it here. In the end the result is
\beq\label{eq:linbeam}
\sigma=a\pi\int d\sigma_{\textrm{\scriptsize red}}\frac1{|k_\bot|}
							\delta(k^2-m^2),
\eeq
where the ``red'' in $d\sigma_{\textrm{\scriptsize red}}$ stands for
``reduced'' meaning that the factors $1/(k^2\pm i\epsilon)$ that cause
the divergence have to be left out. $a$ is the transverse size of the beam.
We take this to be given by
$a=\sqrt\pi\,\sigma=\sqrt\pi\cdot10\,\mu\mathrm{m}$, where (for round
beams) $\sigma$ is
the standard deviation in the position of the particles in the beam in
either direction perpendicular to the beam. This was
also done in~\cite{sing}.
Because this part of the cross section is proportional to this size, the
effect is called ``linear beam size effect''. If there are contributions
to the cross section from parts of phase space away from the singularity
they should be added separately to the cross sections. These parts do
not depend on the beam size. Below we will see that such a separation
arises rather naturally from the shape of the graph of $d\sigma/dk^2$.
The part of the cross section that has to do with regions of phase space away
from the singularity will be called the ``regular cross section'', while
the part that comes from regions near the singularity (or singularities)
will be called the ``semi-singular cross section''.

\begin{figure}
\epsfig{file=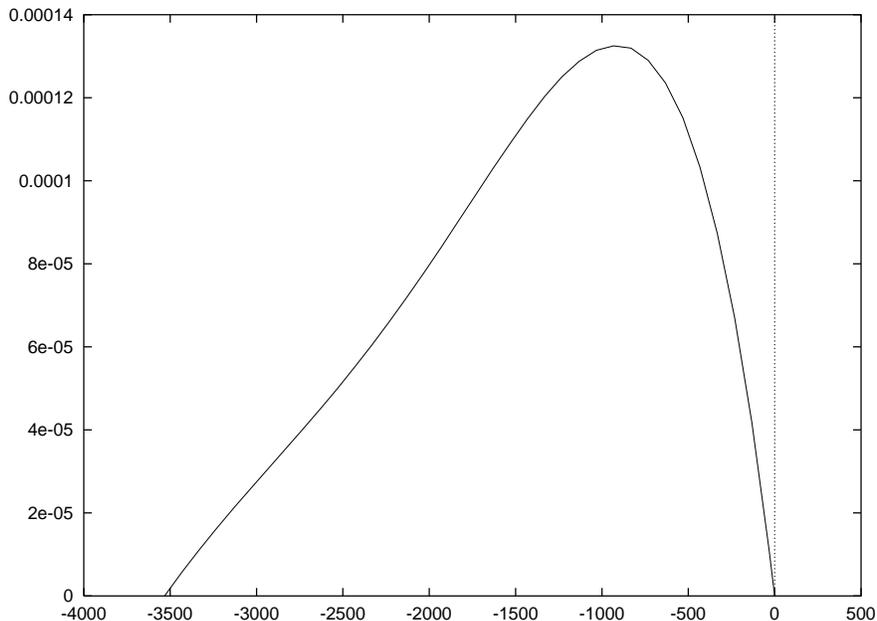,height=\textwidth,angle=270}
\caption{$d\sigma/dt$ (fb/GeV$^2$) vs. $t$ (GeV$^2$) for
$\sqrt s=100\,\textrm{GeV}$}
\label{fig:diff}
\end{figure}

\begin{figure}
\epsfig{file=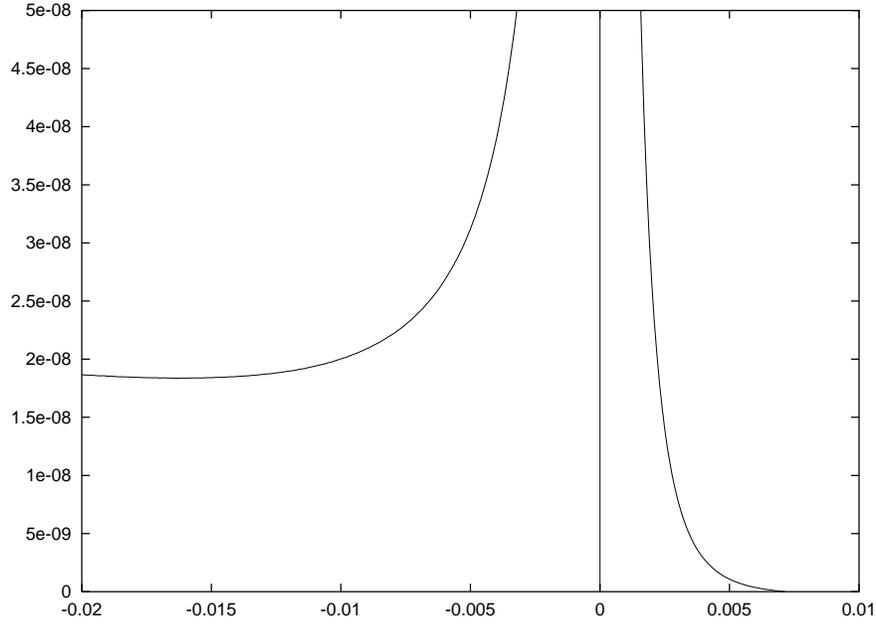,height=\textwidth,angle=270}
\caption{$d\sigma/dt$ (fb/GeV$^2$) vs. $t$ (GeV$^2$) for
	$\sqrt s=100\,\textrm{GeV}$ around $t=0$. This is a detail of the
	previous plot.}
\label{fig:diffdetail}
\end{figure}

\begin{figure}
\epsfig{file=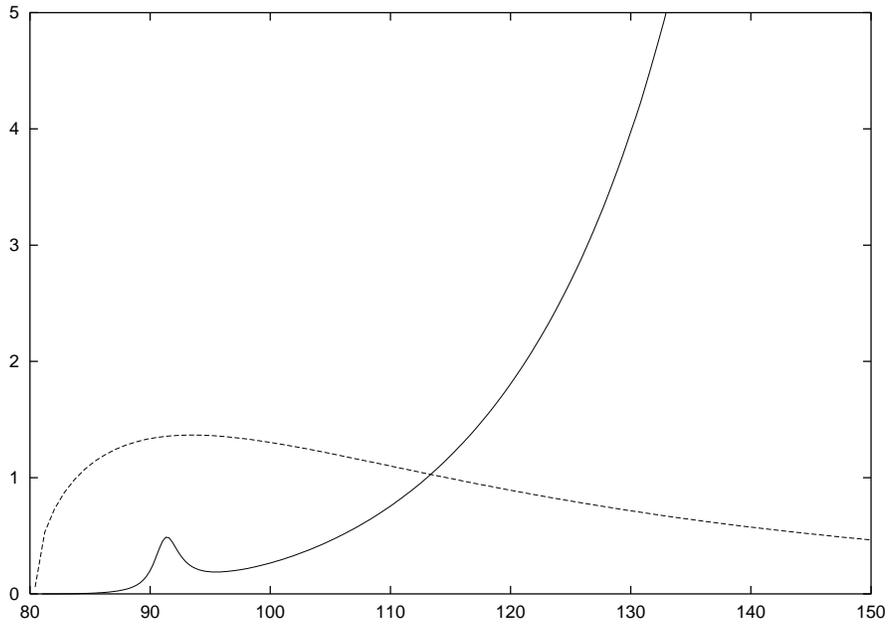,height=\textwidth,angle=270}
\caption{$\sigma$ (fb) vs. $\sqrt s$ (GeV). The solid line is the regular
cross section and the dashed one is the pseudo-singular cross section.
$a=\sqrt\pi\cdot 10\,\mu\mathrm{m}$.}
\label{fig:crosssect}
\end{figure}

As in \cite{sing}, we consider the process
$\mu^-+\mu^+\to W^++e^-+\bar\nu_e$.
For $d\sigma/dt$ at 100\,GeV we find the graph in figure~\ref{fig:diff}.
The infinite spike at~$t=0$ is caused by the instability.
Figure~\ref{fig:diffdetail} shows a detail
of the same graph. Now the singularity is prominently present. The reader
can compare the numbers on the axes of both graphs to get an idea how
narrow the singularity actually is. To be able to calculate the regular
part of the cross section without having to worry about the singularity,
the cut~$t<-m_\mu^2$ is used (as was done in~\cite{sing}).
The two graphs just shown, justify this cut
(i.e., $m_\mu^2\sim 0.01\,\textrm{GeV}^2$ which is in the neighbourhood of
the minimum of $d\sigma/dt$).

The total cross section for pseudo-singular and regular cross sections are
plotted in figure~\ref{fig:crosssect}. For the regular cross section we find
the same graph as in~\cite{sing}, but for the semi-singular cross section
our graph is about a factor~1.7 higher. The reason for this difference
appears to be twofold. In the first place, we did our calculations from
standard model coupling constants, while~\cite{sing} expresses the
cross section in other decay widths and cross sections. If we take this
into account, the factor~1.7 becomes a factor~2. This factor~2 is then
due to an error in equation~46 in~\cite{sing}. This equation should have
an extra factor~2 on the right hand side.

Consequently we find that the semi-singular cross section
dominates up to about 113\,GeV. \cite{sing}~has 105\,GeV.
Furthermore it should be noted that at about
90\,GeV the semi-singular and regular cross sections are about the same
order of magnitude because the regular one has a peak there, caused by the
$Z$~particle. For $\sqrt s$ a bit above threshold the semi-singular cross
section dominates strongly. At about, say,
$\sqrt s\sim 85\,\textrm{GeV}$ one
can safely forget about the regular part. Above $\sqrt s\sim150\,\textrm{GeV}$
the pseudo-singular cross section does not play any role anymore. This is the
case up to arbitrary high energies because asymptotically the semi-singular
cross section goes down as~$1/(s\sqrt s)$, while the regular one goes
down as~$1/s$.

To calculate these cross sections, six diagrams involving
$\gamma$,~$W^\pm$ and~$Z^0$ as fundamental bosons were added.
The algebra necessary was done by the C++ computer algebra
library GiNaC which is described in~\cite{ginac}. After that the integrations
were carried out by adaptive Simpson integration. Unstable intermediate
particles were given propagators using the $iM\Gamma$ prescription. This,
of course, raises the issue of gauge invariance. It was checked that
for high energy the cross section goes down as $1/s$, but there may still
be unnoticed terms suppressed by a factor $\Gamma^2/M^2$ that did not
show up in these calculations. The authors hope to address these
issues in a subsequent paper.

In practice one does not need the cut-off $t<-m_\mu^2$. This is because
this cut-off is already implied by the cut-offs imposed by measurability.
If we take $\sqrt s=100\,$GeV and demand that
the energy of the outgoing electron of the pseudo-singular process is
at least $1\,$GeV and that the angle under which this electron
appears is at least $2^\circ$ we have made it impossible for the
muon neutrino to be on its mass shell (actually we then find that $t$
indeed is always negative and has an upper limit of about $-5m_\mu^2$).

However, if one wants to include the linear beam size effect in the
normal Monte Carlo integration procedure, one can do this by doing the
replacement
\beq
\frac i{k^2+i\epsilon}\to\frac i{k^2+i|k_\bot|/a}.
\eeq
This, in spite of its ad-hoc appearance, gives precisely the correct
answer.
It effectively introduces a decay to the muon neutrino to account for the
fact that it can no longer collide after it has left the beam. This way of
handling the divergence supposes that number
of particles in the beam goes to zero
exponentially as we leave the beam
(i.e., the $iM\Gamma$~prescription is about exponential decay),
which is not terribly realistic, but
in most cases this will not matter.

\section{Infinitely Wide Beams}
\subsection{The Wrong Way}
The method of calculation introduced in~\cite{ginz} (to be called Ginzburgs
method in the rest of this section) starts by observing
that the propagator of an unstable particle is given by
$(p^2-m^2+im\Gamma)^{-1}$. We then ``conclude'' that the mass of the unstable
particle has acquired an imaginary part. The four momentum-squared of this
particle should be complex too. In its rest frame it is given by
\beq
p_1=(m-i\Gamma/2,\vec 0).
\eeq
We now write for the ``new'' value of $k^2$, that is the value that takes
the complex momentum components into account,
\beq
k^2_{\textrm{\scriptsize new}}=(p_1-q)^2=m^2-im\Gamma+q^2-2(m-i\Gamma)q^0,
\eeq
where $q=q_1+q_2$ as drawn in equation~\ref{eq:fmdiag}.
For some reason we take the same values for the components of~$q$ but only
change the components of $p_1$~and $k$. Normally (i.e., without complex
momentum components), the value of $q^0$
is given by
\beq
q^0=\frac{m^2+q^2-k^2}{2m}.
\eeq
After substituting this, and neglecting the small quantity proportional to
$k^2\Gamma$, we obtain
\beq
k^2_{\textrm{\scriptsize new}}
	=k^2_{\textrm{\scriptsize old}}-i\frac\Gamma{2m}(m^2-q^2).
\eeq
This complex value then replaces the one given 
in equation~\ref{eq:prop}, and a finite result is obtained.

The problem with all this is, of course, that it is not terribly difficult to
think of a process where one of the outgoing/incoming particles involved is
unstable and then an incoming complex momentum component has, by momentum
conservation, no place to go.
Significantly, we never hear about momentum conservation for the other
vertex in the diagram.

In~\cite{phys} it is shown that the result obtained by using the here
described method is exactly what one would expect for a muon that decays
in a medium of anti-muons.
This result is
\beq\label{eq:split}
\sigma=\int (1-\cos\theta) w(\omega)\,d\omega
	\frac{\sin\theta\,d\theta\,d\phi}{4\pi}
	\sigma_{\nu\mu\to W}(s_{\nu\mu}).
\eeq
The factor $w(\omega)\,d\omega\,\sin\theta\,d\theta\,d\phi/4\pi$ is the
probability measure of finding a muon neutrino with a certain momentum.
This calculation is done in the rest frame of the decaying muon.
However, three remarks are in order here\footnote{An email discussion
with V.G. Serbo was helpfull to get these points completely clear.}
\items{cijfers}{9}
\item To obtain this result, the definition of the quantity cross section
has to be modified;
\item The same modification of definition can be aplied to our
result (i.e., equation~\ref{eq:result}, to be derived in the next section)
and the result will be the same;
\item It is a bit of a coincidence that the modified cross section
of~\cite{phys} turns out to have the same value as Ginzburgs method
\einditems
Let us discuss these points in this order.

Firstly \textbf{(1)},
the number of events~$W$ is related to the cross section via
\beq
W=V_4\sqrt{(J_1\cdot J_2)^2-J_1^2J_2^2}\,\sigma_{\mu\mu},
\eeq
where $V_4$ is the four-volume in which the beams overlap and $J_{1,2}$ denotes
the four-flux of the two beams. The space integral over $J_1$ is the
number-of-particles four-vector~$N_1^\mu$, defined by
\beq
N^\mu=Nu^\mu,
\eeq
with $N$ the number of particles. We thus have
\beq\label{eq:num1}
\frac{dW}{dt}=\sqrt{(N_1\cdot J_2)^2-N_1^2J_2^2}\,\sigma_{\mu\mu}.
\eeq
On the other hand, we expect to be able to calculate the number of
events from considering collisions between muon neutrino's and muons,
taking the momentum distribution of the muon neutrino's into account.
Doing so we find
\beq\label{eq:num2}
\frac{dW}{dt}=\int w(\omega)\,d\omega
        \frac{\sin\theta\,d\theta\,d\phi}{4\pi}
	\sqrt{(N_\nu(\omega)\cdot J_2)^2-N_\nu(\omega)^2J_2^2}
	\,\sigma_{\nu\mu}(s(\omega))
\eeq
After specializing to the rest frame of the $\mu^-$ we see that the
division of the two flux factors gives the $1-\cos\theta$ (For this
one has to assume that $k^2=p_2^2=0$.) so that
we indeed find equation~\ref{eq:split} back. The modification to the
definition of the cross section is that
one equates these two $dW/dt$'s. The first $dt$ refers to the time the
$\mu^-$~track spends in the $\mu^+$~cloud while the second~$dt$ refers
to the time the decay product spends in this cloud. In a $\mu^+$ cloud
of infinite size equating these two indeed would seem
to be the only thing that could
give a finite result. However, one should realize that for any cloud of
particles of finite size the standard definition of the cross section
involves and integral over time and then the quotient of the two just
mentioned times will appear in the result. This will depend on
beam shapes.

Secondly \textbf{(2)}, also our result (i.e., equation~\ref{eq:result}),
to be derived
in the next section, contains a factor $d^4r$. One could also pull a
factor~$dt$ out of this and obtain exactly the same result as by
Ginzburgs method.
So, whether or not one likes the equating of $dW/dt$'s mentioned in the
foregoing point, one does not need complex momentum conservation to
obtain the result that belongs to it.

Thirdly \textbf{(3)}, the matrix element, as seen in position space, used
in Ginzburgs method, is not as advertised in~\cite{phys}. To see this,
we should realize that
the method of Ginzburg only modifies the propagator of the muon neutrino.
In particular, nothing is altered in the prescription of removing external
propagators. The consequence of this is that the number of muons does not
decrease. After all, this prescription was designed to describe a stable
particle that comes in from infinity. This means that the production rate
of neutrino's inside the $\mu^+$ cloud is given (in the
rest frame of the $\mu^-$) by
\beq
\frac{dN_{\nu_\mu}}{dt}=N_\mu\Gamma_\mu,
\eeq
Thus, every muon produces a large number of neutrino's. Because of the
conservation of complex momenta, the muon neutrino gets a decay time equal
to the decay time one would expect for the muon. The consequence is that
the number of muon neutrino's at a particular time is given by
\beq
N_{\nu_\mu}(t)
	=\int dt'\,\theta(t-t')\,e^{-\Gamma_\mu(t-t')}\frac{dN_{\nu_\mu}}{dt'}
	=N_\mu.
\eeq
So in the method of Ginzburg the number of neutrino's is equal to the
number of muons not because every muon decays into a single muon neutrino
but because the decay constant of the muon neutrino is artificially made
equal to its production constant. That this gives the same result as
equating the two $dW/dt$'s as discussed under point~\textbf{(1)} can be
easily understood because the muon neutrino's now only exist near the
path of the $\mu^-$, so the two $dt$'s now both refer to the time the
$\mu^-$ spends in the $\mu^+$ cloud.

\subsection{The Right Way}
Here we take the production process of the unstable particle into account.
This is reasonable because unstable particles cannot really be asymptotic
states. In the context of scalar fields, Veltman (cf.,~\cite{veltman})
has shown that if one takes only stable particles for asymptotic
states, one gets a unitary S-matrix.
Thus we must describe the unstable particle as an internal line of
a larger Feynman graph that includes the production process.
Consequently its momentum can have
four independent components. Of course, we really do not want to include much
information about the production process of the unstable particle in our
calculations. For this reason we define the wave function of the unstable
particle to be given by
\beq
\psi(p)=\sqrt{\frac2{m_1\Gamma}}
	\int\tie{p_a}\,\tie{p_b}\,\phi_a(p_a)\phi_b(p_b)\,
        (2\pi)^4\delta^4(p_a+p_b-p)\,\Mel(p_a,p_b,p),
\eeq
where $\Mel$ is the matrix element (or perhaps the sum of matrix elements)
that describe the production process.
This definition assumes that the unstable particle in produced in a two-to-one
collision, however we can safely include terms where the production process
has as many in- or outgoing particles as we would like. The outgoing particles
are represented by complex conjugates of the wave functions that they are
measured to be in. Such a measurement should be carried out in order to make
sure that the momentum of the unstable particle is fixed very accurately.
When doing a
calculation with this we still have to include the propagator of the unstable
particle. The just given wave function only replaces the part of the matrix
element that describes the production process, not the propagation of the
unstable particle. For this propagator, the replacement
\beq
\frac i{p^2-m^2+im\Gamma}\to\frac1{m\Gamma}
\eeq
should be used, since the momentum of the unstable
particle is assumed to be
fixed---by the just-described measurement procedure---on its mass shell
so accurately that a function of which the width is of the order~$\Gamma$
(as the propagator is) cannot notice the difference.

The normalization of this wave function, as occurs in its definition above,
was obtained by considering the
production process followed by the decay process. The number of events in
this is demanded to be unity because we choose our wave function to describe
one unstable particle.

In this case the definition of the quantum distribution function should
be modified a bit. It becomes
\beq\label{eq:qdist}
n(p,r)=\int\frac{d^4\Delta}{(2\pi)^4}e^{-i\Delta\cdot r}
        \psi(p+\tfrac12\Delta)\psi(p-\tfrac12\Delta)^*.
\eeq
Probability densities in momentum and position space can be obtained
from this. They are given by
\beq
\rho(p)=\int d^4r\,n(p,r);\qquad \rho(r)=\int\frac{d^4p}{(2\pi)^4}\,n(p,r).
\eeq

The luminosity is now defined by
\beq
dL(p_1,p_2,\rho)=
        \frac{\sqrt{-\mathrm{GD}(p_1,p_2)}}{m_1m_2\Gamma}
        \frac{d^4p_1}{(2\pi)^4}\tie{p_2}\int d^4 r\,n_1(p_1,r)
        n_2(p_2,r+\rho).
\eeq
This is for the case of a stable particle colliding with an unstable one.
The reader will presumably have no difficulty figuring out what to use for
two unstable particles if he compares this to equation~\ref{eq:stablum}.
The cross section is defined in exactly the same way as in
section~\ref{sec:realistic}.
Furthermore, the just given definition for
the luminosity was chosen in such a way that equation~\ref{eq:numevents}
is kept the same too.

After having introduced this, the calculation of the number of events
proceeds
along exactly the same lines as in~\cite{sing} and we will not present
it here. The result is
\beq\label{eq:result}
\eqalign{
W=\frac1{m_2}\int\frac{d^4 p_1}{(2\pi)^4}\tie{p_2}\,
        \tie{k}\,d^4r\int_0^\infty d\alpha\,
        n_1(p_1,r)n_2(p_2,r+\alpha k)f_s(k|p_1)\qquad\cr
        \sqrt{(k\cdot p_2)^2-k^2p_2^2}\,\sigma(k,p_2).
}
\eeq
The interpretation of the above formula is that the unstable particle decays
at position~$r$, the decay product has momentum~$k$ and then collides with
the other particle at position~$r+\alpha k$, where $\alpha$ is a positive
number, as one would expect.
If we specialize to the case where the longitudinal beam size is much larger
than the transversal one, we find back formula~\ref{eq:linbeam}.
The densities that occur here should be considered to be densities of
decay events. It is perfectly natural that decay events are characterized by
four co-ordinates.

\section{Pancake-shaped Beams}
In this section we will show that it is possible to transfer some
of the $im\Gamma$ to the muon neutrino propagator while keeping a clear
conscience. This is in the context op pancake-shaped beams. In this case, 
we assume the transversal beam size to be much larger than the decay
length~$1/\Gamma$, while the longitudinal beam size is much smaller than
the decay length. We do this by using momenta that have a large width
in the longitudinal direction and a small width in the transversal direction.
We assume that the unstable particle is produced at a position away from
the position where it collides. We do this by translating the particle
having momentum~$p_2$ in equation~\ref{eq:fmdiag} an invariant distance~$s$
(as seen by the $p_1$-particle) away from the origin.
This is done by a adding factor $e^{isp_1\cdot p_2'/m_1}$ in the matrix
element. The complex conjugated matrix element gets a factor
$e^{-isp_1\cdot p_2''/m_1}$. The variables $p_2'$ and $p_2''$ are integration
variables in momentum space. They integrate over the momentum distribution.

Furthermore, we take into account (as we have done for the case of infinitely
wide beams) the fact
that unstable particles cannot really be
in/out states. The real matrix element describing
the collision process includes
the production of the unstable particle. The incoming unstable particle
differs from a stable one in that its momentum need not be on its mass
shell. The
momentum distribution is determined by its
propagator~$((p_1')^2-m_1^2+im_1\Gamma)^{-1}$, which makes sure that the
momentum is peaked around the mass shell.

We consider the quantity~$F$ which contains the factors of the matrix element
that play a role in making the cross section finite. These are the factors
that are peaked sharp enough to notice that~$p_1'$ (and~$p_2'$, of course)
do not have a
definite value but are peaked around the value~$p_1$.
It is given by
\beq\label{eq:Fdef}
F=\frac{e^{isp_2'\cdot p_1/m_1}}{(p_1')^2-m_1^2+im_1\Gamma}
	\frac1{(q_2-p_2')^2-M^2+i\epsilon}
\eeq
We now integrate over the value of $(p_1')^2$. The integration path is chosen
such that outgoing momenta are kept fixed, while the change~$\Delta^\mu$
of the incoming
momenta is taken to be a linear combination of these momenta. Thus momenta
are parameterized by
\beq
\eqalign{
(p_1')^\mu(t)&=(p_1')^\mu(0)+t\Delta^\mu;\cr
(p_2')^\mu(t)&=(p_2')^\mu(0)-t\Delta^\mu;\cr
(k')^\mu(t)&=(k')^\mu(0)+t\Delta^\mu,\cr
}
\eeq
where $t$ is just a parameter and has nothing to do with the $t$-channel.
Furthermore we should demand that the value of $(p_2')^2$ is kept
fixed. It is not very difficult to think of a scatter setup where the
particle with momentum~$p_2$ is a stable one, so we had indeed better
not vary this one.
These demands are satisfied by taking
\beq
\Delta^\mu=\frac12
        \frac{(p_1\cdot p_2)p_2^\mu-p_2^2p_1^\mu}{(p_1\cdot p_2)^2-p_1^2p_2^2},
\eeq
We take a linear combination of $p_1$ and $p_2$ because of the pancake
shape. These momenta only have a sufficiently large spread in momentum space
in the longitudinal
direction, so this direction should be chosen for the integration path.
Filling this into equation~\ref{eq:Fdef} and using contour integration over~$t$,
we find that for the matrix element the integration boils
down to the substitution
\beq
\eqalign{
F\to&-2\pi i\frac{e^{isp_2\cdot p_1/m_1-s\Gamma/2}}{k^2-M^2-i\alpha m_1\Gamma}
                \delta(p_1^2-m_1^2)\cr
        &-2\pi i\frac{e^{isp_2\cdot p_1/m_1}}{
                        p_1^2-m_1^2+im_1\Gamma}\theta(\alpha)
                \delta\left(\alpha^{-1}(k^2-M^2)\right),\cr
}
\eeq
where $\alpha=2\Delta\cdot q_2$.

We see that this no longer causes a divergence, so we will square
this and work out the Golden Rule to find a cross section.
The square of both terms contains a Breit-Wigner of which the square
can be approximated by a $\delta$~function.
However, before we do this, we should discuss the meaning of the two
terms.
In the first one we effectively first integrate over the width of the
unstable particle and then over the width of the decay product. In the
second term it is the other way around. The interpretation is that the
width that is integrated over first corresponds to the smallest virtuality
(this corresponds to the largest distance scale).
Thus, the first term describes events where the unstable particle
does not decay or decays nearby the spot where the collision happens.
The second term
describes events where the unstable particle decays long before the decay
product collides. We interpret this as the case where the unstable particle
is far off shell and is never really produced but already decays during
the production process.
We decide to drop the second term and keep the first one.
If we square the term that we decided to keep, we get the familiar decay
law (i.e., the $e^{-\Gamma s}$).

The consequence of the just described procedure is that our squared
matrix element contains a factor
\beq
\frac1{(k^2-M^2)^2+\alpha^2m_1^2\Gamma_1^2}\sim\frac\pi{\alpha m_1\Gamma}
								\delta(k^2-M^2).
\eeq
For the cross section, we find
\beq
\sigma(p_1,p_2)=\int\tie k\,
	\frac{\sqrt{(p_2\cdot k)^2-p_2^2k^2}}{\sqrt{(p_1\cdot p_2)^2-p_1^2p_2^2}}
	\frac{e^{-s\Gamma}}\alpha\sigma(p_2,k)f_s(k|p_1).
\eeq
The quotient of flux factors is the generalization of the
$1-\cos\theta$ in equation~\ref{eq:split}. The only factor that we might not
have expected is the $1/\alpha$.

We can shed light on this factor~$1/\alpha$ in the rest frame of the stable
particle (the one with momentum~$p_2$). In that frame we have, writing out
all inner products,
\beq
\sigma(p_1,p_2)=\int\tie k\,\frac{e^{-s\Gamma}}{|\cos\theta|}\sigma(p_2,k)f_s(k|p_1).
\eeq
We see that the flux factor together with the $1/\alpha$ turn into a factor
$1/\cos\theta$. This is because
we are
describing the collision of two pancake-shaped beams. The $p_2$~particle
is taken to be in its rest frame here, so it should be imagined to be a
pancake that does not move. The decay product that collides with a particle
in the pancake emerges under an angle~$\theta$. Because of this, it sees the
thickness of the pancake enlarged by a factor $1/\cos\theta$.

\section{Conclusions}
The problem that the cross section diverges for colliding unstable particles
can be considered for the case of beams much smaller than the decay length
or for beams much larger than the decay length. We also had a look at
what happens if two pancake-shaped beams collide.

In the first case, the divergent part of the cross section is proportional
to the transverse beam size. We can confirm the result of~\cite{sing}
that for $\sqrt s\approx 100\,$GeV the part of the cross section independent
of the beam size and the part proportional to it are of the same order of
magnitude. However, for higher energies ($\sqrt s> 150\,$GeV already)
the part proportional to the beam size is at least three orders of magnitude
smaller. At still higher energies this becomes even more, so for high
energy colliders the linear beam size effect can safely be ignored. In
practice this can be done by imposing cuts as one would have in a
collider.

In the second case, it
should not be ``solved'' by introducing complex momentum components. 
This artifically introduces a decay time for the decay product, which does
not seem to be a real physical effect. The authors of this paper are not
aware of the existence of a law of ``conservation of decay width''.
The result that is obtained by the introduction of complex momentum
co-ordinates can (if we allow for a not unreasonable modification in the
definition of the cross section) also be obtained by using our methods.
Our methods involve considering the wave function of the
unstable particle to be a function of four momentum components. This is
reasonable because the unstable particle cannot be an in/out state and
really is an internal line of a bigger Feynman diagram. After this, the
reasoning proceeds along the same lines as for the case of realistic beams.
We find (if we do not allow for the just mentioned modification and assume
cylindrically shaped beams) the same linear beam size effect.

Another method, namely
integrating over the width of $p^2$ of the unstable particle, is the right
thing to do for the case of pancake-shaped wave packets.

One might ask to what extent delicate gauge cancellations are destroyed
by our approach. We do not expect these problems to be qualitatively
worse than those encountered in, say, calculating loop corrections to
LEP-2 processes (cf.~\cite{beenakk}). This point will be addressed in a
forthcoming publication.

\end{document}